\documentstyle[aps,amsfonts,epsfig,psfig]{revtex}

\begin{document}

\newlength{\insideFig}

\title{Atomic Energy Levels with QED\\ and Contribution of the Screened
  Self-Energy}

\author{\' Eric-Olivier Le~Bigot, Paul Indelicato
}

\address{
Laboratoire Kastler-Brossel,
\' Ecole Normale Sup\' erieure et Universit\' e P. et M. Curie\\
Unit\' e Mixte de Recherche du CNRS n$^\circ$ C8552, Case 74\\
4, pl.~Jussieu, 75252 Paris CEDEX 05, France%
}

\maketitle

\begin{abstract}
  We present an introduction to the principles behind atomic energy
  level calculations with Quantum Electrodynamics (QED) and the
  two-time Green's function method; this method allows one to calculate
  an effective Hamiltonian that contains all QED effects and that can
  be used to predict QED Lamb shifts of degenerate, quasidegenerate and
  isolated atomic levels.
\end{abstract}

\section*{Introduction}

This contribution is concerned with the evaluation of atomic energy
levels with QED. Such an evaluation yields stringent tests of QED in
\emph{strong electric fields}, whereas $g$-factor experiments and
calculations currently probe QED in situations where the magnetic
field can be treated perturbatively

The nuclear Coulomb field experienced by the inner levels of
\emph{highly-charged ions} makes the electrons reach
\emph{relativistic} velocities.  Such simple physical systems are
thus particularly interesting for testing relativistic effects in
quantum systems (for example, see
Refs.~\cite{beiersdorfer98,schweppe91} for experimental results with
lithiumlike ions). Theoretical predictions of energy levels in such
systems obviously \emph{require} the use of QED.

Experiments have reached an accuracy that shows that extremely
accurate evaluations of QED effects are also needed in
\emph{helium}.  Experiments performed during the last ten years
in the spectroscopy of this
atom
have become two orders of magnitude more precise than the current
theoretical calculations (see for instance
Refs.~\cite{drake98,drake2000} and references therein).

Several experiments are now focusing on helium and heliumlike ions,
and especially their $1s2p\,^3\! P_J$ fine
structure~\cite{minardi99,storry2000,castillega2000,myers2000}
such experiments have implications in metrology, as they could provide
a measurement the fine structure constant and provide checks of
theoretical higher-order effects. Very precise theoretical
calculations of energy levels in heliumlike ions can be also important
in the investigation of \emph{parity
  violation}~\cite{maul96}.

Predictions of energy levels are usually more difficult to obtain for
states with one or more open shells (retardation in the interaction
and exchange of electrons must be included, and there can be
quasidegenerate levels). Only a few calculations of excited energy
levels in heliumlike and lithiumlike ions have been performed up to
now; the first results have been published quite
recently~\cite{artemyev99,artemyev2000,mohr2000}. In regards to QED
shifts of \emph{quasidegenerate} levels, they have only been
obtained this year for the first time~\cite{artemyev2000}, with the
help of the method that we present in this talk.

\section*{Theoretical methods}

As is well known, relativistic electrons orbiting a nucleus are well
treated with the \emph{Dirac equation}, in which the nucleus can
be considered as point-like or not. We thus treat the binding to the
nucleus \emph{non-perturbatively} by using ``Bound-State
QED''~\cite{furry51,mohr89} (the coupling constant of the
nucleus-electron interaction is $Z\alpha$, which is not small for
highly-charged ions). In this formalism, however, QED effects are
taken into account by treating the electron-electron interaction
\emph{perturbatively} (with coupling constant $\alpha$), and both
the electron and photon fields are \emph{quantum} fields (i.e., in
\emph{second}-quantized form); the only difference with the
free-field case used in high-energy physics is that electronic
creation and annihilation operators create and destroy atomic states
instead of free particles.

A few methods allow one to extract energy levels from the Bound-State
QED Hamiltonian: the two-time Green's function
method~\cite{shabaev94,shabaev94b,shabaev2000}, the method being
developed by Lindgren (based on Relativistic Many-Body Perturbation
Theory merged with QED)~\cite{lindgren2000,specialLindgren2001}, the
adiabatic $S$-matrix formalism of Gell-Mann, Low and
Sucher~\cite{sucher57}, and the evolution operator
method~\cite{vasilev75,zapryagaev85}. Some other methods yield atomic
energy levels, but they include QED effects only \emph{partly} or
approximately (such as the multiconfiguration Dirac-Fock
method~\cite{indelicato90}, configuration interaction
calculations~\cite{cheng2000} and relativistic many-body perturbation
theory~\cite{ynnerman94}).

However, only \emph{two}  methods can in principle be
employed in order to calculate energy levels of
\emph{quasidegenerate} atomic states [e.g., the $(2s2p_{1/2})_1$
and the $(2s2p_{3/2})_1$ levels in heliumlike ions, which are
experimentally important]: the two-time Green's function method and
the method being elaborated by Lindgren. We present in this talk a
non-technical introduction to the first method. The two-time Green's
function method has also the advantage of yielding a
\emph{simpler renormalization procedure} than the
Gell-Mann--Low--Sucher method in the case of \emph{degenerate}
levels~\cite{braun80,braun84}.

\section*{The two-time Green's function method}

All the methods that extract atomic energy levels from the Bound-State
QED Hamiltonian study the \emph{propagation} of electrons between
two different times. The methods differ in the number of
\emph{infinite} times used: 

(a)~in the Gell-Mann--Low--Sucher method, the atomic state under
  consideration evolves from time $-\infty$ to time $+\infty$ with an
  adiabatic switching of the interaction;
(b)~in Lindgren's formalism~\cite{lindgren2000,specialLindgren2001},
  the evolution is from time $-\infty$ to time $0$, which avoids
  problems associated with the two infinite times in the $S$-matrix
  approach of Gell-Mann--Low--Sucher;
(c)~in the two-time Green's function method, that we present here,
  the adiabatic switching is completely avoided by studying the
  propagation of electrons between \emph{two finite times}.
We note that adiabatic switching of the interactions is physically
motivated in the study of collisions between particles that start very
far from each other, but this switching is not so easily related to
the physical description of the orbiting electrons of an atom.

\subsection*{The Green's function}

\begin{figure}[!htb]

  \begin{center} 
  {\resizebox{0.5\textwidth}{!}
      {\epsfig{file=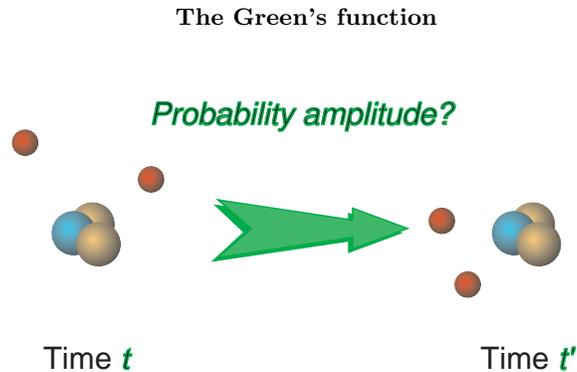}}}
  
  \end{center}
  
\vspace{10pt}
\caption{The two-time Green's function represents the probability
amplitude for going from one position of the electrons to another position.}
\label{fig:green}
\end{figure}

The \emph{effective Hamiltonian} derived from QED by the
  two-time Green's function method has matrix elements between the
  various degenerate and/or quasidegenerate states under study; the
eigenvalues of this Hamiltonian are the atomic energy levels predicted
by QED (to a given order). This effective Hamiltonian is however
\emph{not} associated to a Schr\" odinger equation of motion; our
Hamiltonian is equivalent to the submatrix used in the perturbation
theory of degenerate and quasidegenerate states; in this respect, the
approach of the two-time Green's function method differs from the
spirit of the Bethe-Salpeter equation.

The QED Hamiltonian of the method is defined with the help of a
Green's function that represents the propagation of $N$ electrons
between two different (finite) times ($N$ is the number of electrons
of the atom or ion that we want to study); this propagation is
represented in Fig.~\ref{fig:green}.

\subsection*{Atomic energies are in the Green's function}

\begin{figure}[!htb]

  \begin{center} 
  {\resizebox{0.5\textwidth}{!}
      {\epsfig{file=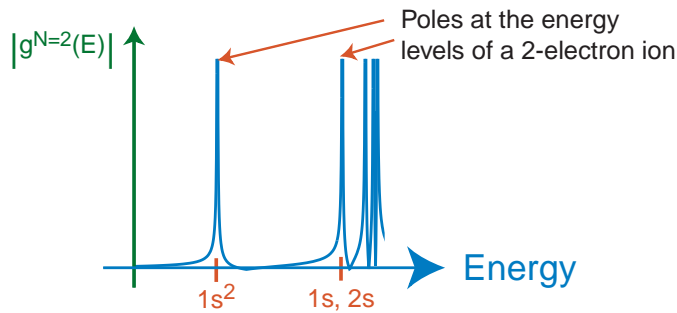}}}
  
  \end{center}
  
\vspace{10pt}
\caption{The two-particle Green's function as a function of
  \emph{energy} contains all the information about the atomic
  energy levels of a two-electron atom or ion.
}
\label{fig:greenPoles}
\end{figure}

The energy levels of an $N$-electron ion or atom can be recovered by
studying the \emph{energy} representation ${\mathcal{G}}^{N}(E)$ of
the Green's function%
, i.e., by doing a Fourier transform:
this function \emph{has (simple) poles at the atomic energy
  levels}~\cite{shabaev94,shabaev94b,shabaev2000}.
Such a result is
similar to the K\" all\' en-Lehmann representation~\cite{peskin95}.
As an example, Fig.~\ref{fig:greenPoles} depicts the poles of the two-particle
Green's function.

\begin{figure}[!htb]

  \begin{center} 
  {\resizebox{0.5\textwidth}{!}
      {\epsfig{file=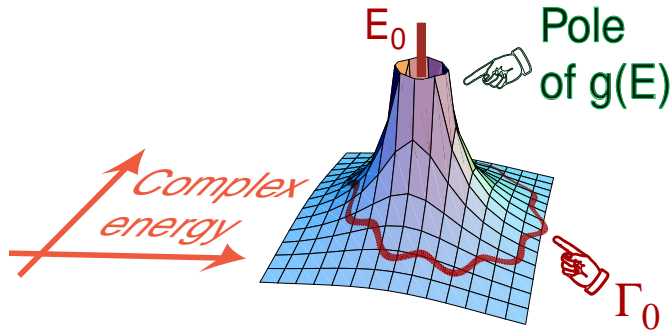}}}
  
  \end{center}
  
\vspace{10pt}
\caption{The exact atomic energies can be recovered through a contour
  integration of the Green's function.
}
\label{fig:contourInt}
\end{figure}

The two-time Green's function method provides a way of
\emph{mathematically} extracting from the Green's function the
positions of the poles, i.e., the atomic energy
levels~\cite{shabaev2000}; the procedure handles degenerate and
quasidegenerate atomic levels without any special
difficulty~\cite{shabaev93}. One of the basic ideas behind the pole
extraction is found in the following mathematical device, which uses
\emph{any} contour $\Gamma_0$ that encloses the pole in order
to find its exact position: if the function $g(E)$ has a simple pole
at $E=E_0$, then we have from complex analysis
\begin{equation}\label{eq:poleRecover}
E_0 = \frac{\displaystyle {\displaystyle \oint_{\Gamma_0}\!}
  d{E}\, 
E\times g(E)}
{{\displaystyle \oint_{\Gamma_0}\!}
  d{E}\, g(E)};
\end{equation}
the contour $\Gamma_0$ is only required to encircle the pole and to be
positively oriented, as shown in Fig.~\ref{fig:contourInt}. Since the Green's
function has simple poles at the atomic energy
levels~\cite{shabaev2000}, Eq.\ (\ref{eq:poleRecover}) is a way of obtaining them.

When QED shifts of degenerate of quasidegenerate levels are
calculated, the \emph{scalar} Green's function $g$ of
Eq.\ (\ref{eq:poleRecover}) is simply replaced by a finite-size \emph{matrix}
defined on the space of levels under consideration~\cite{shabaev93}.

\section*{Graphical calculations}

Obviously, \emph{analytic} properties of the Green's
function~\cite{braun84} are important in the evaluation of
Eq.\ (\ref{eq:poleRecover}). We have developed a set of \emph{graphical}
techniques that allow one to obtain the Laurent series of the Green's
function ${\mathcal{G}}^{N}(E)$ by a \emph{systematic} procedure. The
idea behind these techniques consists in displaying the analytic
structure of the Green's function step by step; each step explicitly
extracts \emph{one} singularity, and we proceed until we have
exhausted all the singularities of the Green's function; at this
point, contour integrals such as Eq.\ (\ref{eq:poleRecover}) can be calculated
quite simply.

It is impossible to give here a full account of the method we use for
deriving the effective, finite-size QED Hamiltonian. However, we can
mention a particular feature of our calculational strategy: a very
special ``particle'' appears in our algorithm; this particle is quite
simple since it ``disintegrates'' immediately (zero life time) and
cannot move (zero probability for going from one position to a
different one). In mathematical terms, the coordinate-space propagator
of this particle is a four-dimensional Delta function $\delta^{(4)}[
(\vec{x}, t); (\vec{x'}, t')]$ that we represent by a special
  line in Feynman diagrams.

\section*{The screened self-energy}

The experimental accuracy on transition energies is so high that
second-order (i.e., two-photon) effects must be taken into account in
order to compare experiments with theory.  We thus have very recently
calculated the contribution of the self-energy
screening~\cite{lebigot2000} to the QED effective hamiltonian; this
contribution corresponds to  the following physical processes:
\[
\parbox{35mm}{ %

\psfig{figure=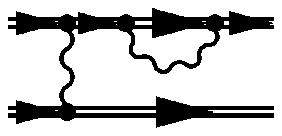}

}
\qquad
\parbox{35mm}{ %

\psfig{figure=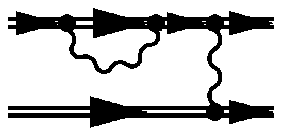}

}
\qquad
\parbox{35mm}{ %

\psfig{figure=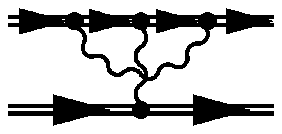}
.
}
\]
Our result is part of the current theoretical effort developed with
the aim of matching experimental precisions.

\end{document}